# Atomic and electronic structure of graphene oxide/Cu interface

Danil W. Boukhvalov[1,2], Ernst Z. Kurmaev[2,3], Ewelina Urbańczyk[4], Grzegorz Dercz[5], Agnieszka Stolarczyk[4], Wojciech Simka[4], Andrey I. Kukharenko[2], Ivan S. Zhidkov[2], Anatoly I. Slesarev[2], Anatoly F. Zatsepin[2], and Seif O. Cholakh[2]

[1]*Department of Chemistry, Hanyang University, 17 Haengdang-dong, Seongdong-gu, Seoul 04763, Korea*

[2] *Institute of Physics and Technology, Ural Federal University, Mira Str. 19, 620002 Yekaterinburg, Russia*

[3]*M.N. Mikheev Institute of Metal Physics of Ural Branch of Russian Academy of Sciences, 18 Kovalevskoj Str., 620990 Yekaterinburg, Russia*

[4]*Faculty of Chemistry, Silesian University of Technology, B. Krzywoustego Street 6, 44-100 Gliwice, Poland*

[5]*Institute of Materials Science, University of Silesia, 75 Pułku Piechoty Street 1 A, 41-500 Chorzów, Poland*

*The results of X-ray photoemission (XPS) and valence bands spectroscopy, optically stimulated electron emission (OSEE) measurements and density functional theory based modeling of graphene oxide (GO) placed on Cu via an electrophoretic deposition (EPD) are reported. The comparison of XPS spectra of EPD prepared GO/Cu composites with those of as prepared GO, strongly reduced GO, pure and oxidized copper demonstrate the partial (until C/O ratio about two) removal of oxygen-containing functional groups from GO simultaneously with the formation of copper oxide-like layers over the metallic substrate. OSEE measurements evidence the presence of copper oxide phase in the systems simultaneously with the absence of contributions from GO with corresponding energy gap. All measurements demonstrate the similarity of the results for different thickness of GO cover of the copper surface. Theoretical modeling demonstrates favorability of migration of oxygen-containing functional groups from GO to the copper substrate only for the case of C/O ratio below two and formation of Cu-O-C bonds between substrate and GO simultaneously with the vanishing of the energy gap in GO layer. Basing on results of experimental measurements and theoretical calculations we suggest the model of atomic structure for Cu/GO interface as Cu/CuO/GO with C/O ratio in gapless GO about two.*

# 1. Introduction

Graphene oxide and its composites are discussed as prospective materials for ink-jet and flexible electronics [1-5] and solar cells. [6-10] These applications require fabrication of interfaces between metallic contacts between GO (graphene oxide). Recent experimental studies demonstrate a significant reduction of graphene oxide in the presence of metallic nanoparticles [11-14] and suggest that migration of oxygen-containing groups (mostly epoxy and hydroxyl groups) with significant changes of chemical composition of both substrate and cover should also be taking into account in the case of formation of GO/metal interfaces. Because copper is widely used industrial material due to excellent thermal and electrical conductivities which provide its extensive application in electrical engineering, microelectronics [15] we choose this material as the most proper for the modeling of GO/metal contacts.

Another application of GO/metal composites is catalysis [16,17] including photocatalysis. [18,19] Understanding of exact atomic structure of composites and effects of interface formation to the electronic structure is the part of the work for further improvement of performance of these composites. Experimentally [20,21] and theoretically [22,23] reported outstanding catalytic properties of Cu/graphene interfaces also suggest for the choosing of copper for the study of these type of composites.

Additional motivation for exploring of carbon/metal interfaces is the development of simple, long living and efficient anticorrosion protective coatings seem to be very important for their use in above applications. Traditionally, chemically reactive metal surfaces are protected against corrosion by the formation of oxide layers, coatings with paints, polymers, organic layers and noble metals or alloys [24]. However, most of these wear- and corrosion-resistant coatings have thicknesses from several to tens micrometers which make them not usable in all applications. It is expected that developing of very thin and transparent protective coatings can solve above problems [25]. Graphene is chemically inert and impermeable to gasses even helium [26-28]. An ideal graphene sheet is composed of only one atomic layer is, thus, the thinnest material that has ever been produced. It is proven that graphene is a good diffusion barrier and can be used as efficient ultrathin anticorrosion coating [29]. Usually, the chemical vapor deposition (CVD) technique is used for deposition of graphene in situ on a copper substrate with the help of hydrocarbon as a carbon source under high vacuum in the presence of catalyst [30]. However, using CVD technique, it is hard to avoid defects in graphene, which often occur on boundaries between different domains [25] which create a galvanic couple between the anodic substrate and cathodic graphene, causing localized corrosion. In connection with this, the development of low-cost, non-vacuum CVD based techniques for synthesis of graphene is very important.

In this study, we proposed an efficient electrophoretic deposition (EPD) method to deposit GO on the surface of copper. EPD is a term for a broad range of industrial processes, including electrophoretic coating or painting, where positively charged particles deposit on the cathode and negatively charged particles deposit on the anode. EPD has recently been attracting attention as a powerful method for the fabrication of nano-structured, thin ceramic composite films on conductive substrates. We have studied X-ray photoelectron spectra of core levels and valence bands and optically stimulated electron emission of GO/Cu composites prepared by the electrophoretic method. Obtained results together with specially performed density functional theory calculations show the fast changes in the atomic structure of GO/Cu composites exposed at ambient atmosphere which is accompanied by the formation of $Cu_2O/CuO$ buffer between Cu substrate and partially reduced GO.

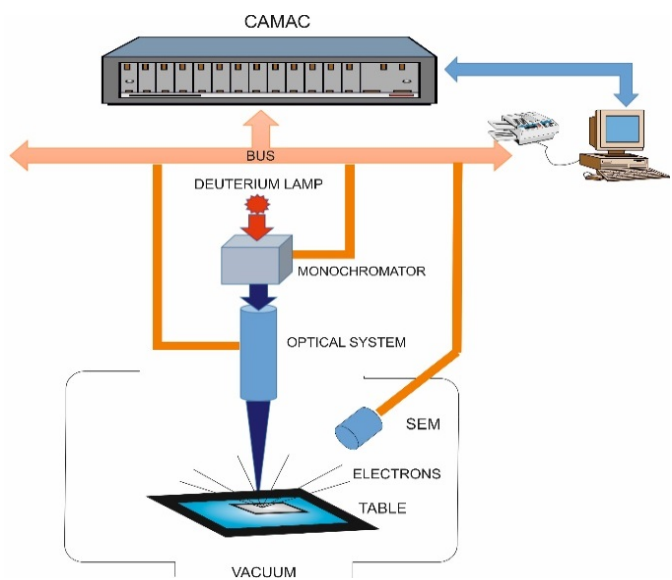

**Figure 1.** The block-scheme of OSEE spectrometer

## 2. Experimental and Calculation Details

The samples used in investigations were made from the copper sheet (M1E Z4, El-Decor, Krakow, Poland) and had dimensions 10x10x0.5 mm. Before the electrophoretic deposition process, the copper samples were ground using #600, #1000, #1500 and #2000 abrasive papers (SiC). The samples were etched in a solution containing 3M $HNO_3$ at 2 min. Then, the samples were subjected to the electrophoretic deposition process. The EPD process was the lead in an aqueous suspension containing 0.2 g $dm^{-3}$ graphene oxide (GO, Graphene Supermarket, USA). Both an anode and a cathode were made of copper. However, the investigated sample was the anode. In this process, the different values of current voltage of 5, 10 and 15 V were used. After the EPD process, the samples were left to dry in air.

The surface morphology of the formed surface layers was examined using a scanning electron microscope (SEM, Quanta 250, FEI). The phase compositions of the layers formed on the copper specimens were determined using a Philips X-Pert Pro PW3040/60 diffractometer operating at 30 mA and 40 kV. A vertical goniometer and Eulerian cradle were used throughout the experiments. The wavelength of the radiation source ($\lambda$=Cu K$\alpha$) was 0.154178 nm. The grazing incidence X-ray diffraction patterns were recorded over the 2$\theta$ range of 10–90° with a step size of 0.05° and an incident angle $\alpha = 0.25°$.

Raman spectroscopy was performed on an inVia Renishaw Raman microscope equipped with a CCD detector using green (514 nm) laser excitation. Scans were taken on a static range (50–3500 cm$^{-1}$). All measurements were made in backscattering geometry, using a 50× microscope objective with a numerical aperture value of 0.75, providing scattering areas of ca 1 $\mu m^2$. Single point spectra were recorded with 4 cm$^{-1}$ resolution and 20 s accumulation times.

X-ray photoelectron spectra (XPS) were measured using a PHI 5000 Versa Probe XPS spectrometer (ULVAC Physical Electronics, USA) based on a classic X-ray optic scheme with a hemispherical quartz monochromator and an energy analyzer working in the range of binding energies from 0 to 1500 eV. Electrostatic focusing and magnetic screening were used to achieve an energy resolution of $\Delta E \leq 0.5$ eV for the Al K$_\alpha$ radiation (1486.6 eV). An ion pump was used to maintain the analytical chamber at $10^{-7}$ Pa, and dual channel neutralization was used to compensate local surface charge generated during the measurements. The XPS spectra were recorded using Al K$_\alpha$ X-ray emission - spot size was 200 µm, the x-ray power delivered at the sample was less than 50 W, and typical signal-to-noise ratios were greater than 10000:3.

The optically stimulated electron emission (OSEE) measurements were carried out with help spectrometer ASED-1 (see Fig. 1). The optical stimulation of the surface of the sample was carried out using the grating monochromator MSD-2, quartz optical system, deuterium lamp DDS-400 and secondary electron multiplier VEU-6 as electron detector. The measurements were performed in the oil-free vacuum of $10^{-4}$ Pa at room temperature.

We used density functional theory (DFT) as implemented in the pseudopotential code SIESTA [31] as in our previous studies (see Ref. 23 and references therein). All calculations were performed using the generalized gradient approximation (GGA-PBE) with spin polarization [32] and implementation of the correction of van der Waals forces. [33] During the optimization, the ion cores were described by norm-conserving nonrelativistic pseudopotentials [34] with cut off radii 1.14, 1.25, 1.20 and 2.15 au for C, N, H, and Cu metals, respectively, and the wave functions were expanded with localized orbitals and double-$\zeta$ basis set for hydrogen and a double-ζ plus polarization basis set for other species. Full optimization of the atomic positions was performed. Optimization of the force and total energy was performed with an accuracy of 0.04 eV/Å and 1 meV, respectively. All calculations were carried out with an energy mesh cutoff of 360 Ry and a k-point mesh of 8 × 6 × 4 in the Monkhorst−Pack scheme.[35]

**3. Results and Discussion**

*3.1. SEM, EDX and TL-XRD investigations of the samples*

The influence of the EPD process voltage on obtained GO coatings mass is presented in the Fig. 2. As seen, during this EPD process the mass of the GO coatings increase with voltage and almost the linear dependence between the mass growth and applied voltage takes place which is similar to that of EPD deposition of GO on carbon mild steel [36]. In the Fig. 3 the SEM and macroscopic images of samples after EPD process are shown. The graphene oxide layers on GO/Cu (5 V) and GO/Cu (15 V) were found to be uniform. Only on the surface of GO/Cu (10 V) sample in some places were brighter areas (see macroscopic image). The color of obtained layers is changed from light brown to dark brown with an applied current voltage which may be caused by the partial reduction of GO or by the increasing thickness of this layer [37,38]. On the SEM images, the graphene oxide layers with increasing the value of voltage are more visible. The morphology of obtained layers is typical for GO coatings on another substrate [39]. A distribution of carbon, oxygen and copper on the samples surface is very homogenous and independent of the EPD process voltage (see an EDX mapping - Fig. 4).

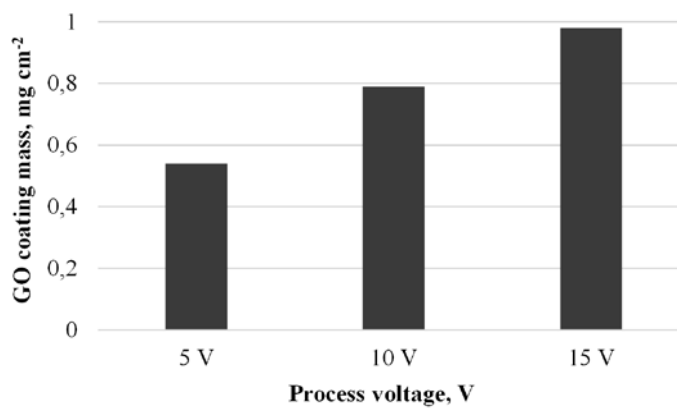

**Figure 2.** Influence of the EPD process voltage on obtained GO coatings mass

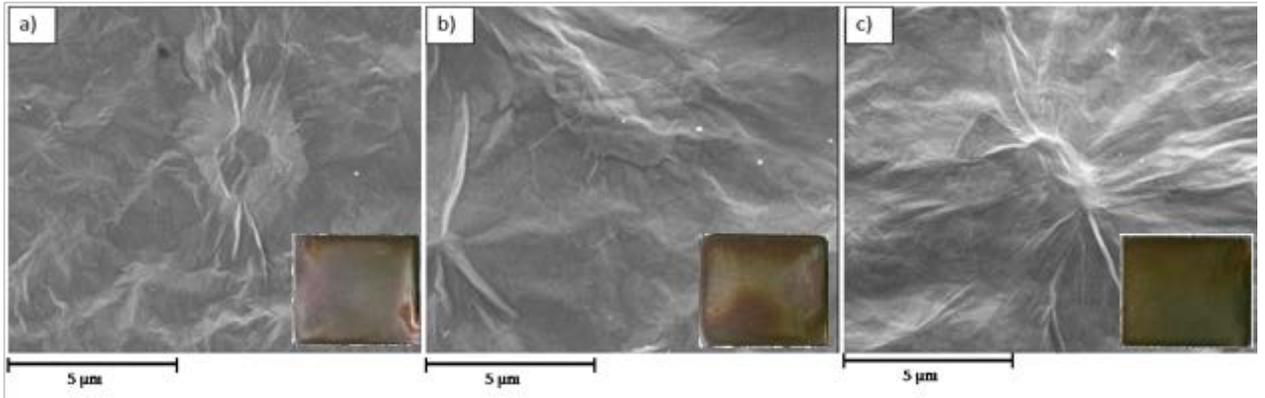

**Figure 3.** The SEM and macroscopic images of copper surfaces after EPD process: a) GO/Cu (5 V), b) GO/Cu (10 V) and c) GO/Cu (15 V)

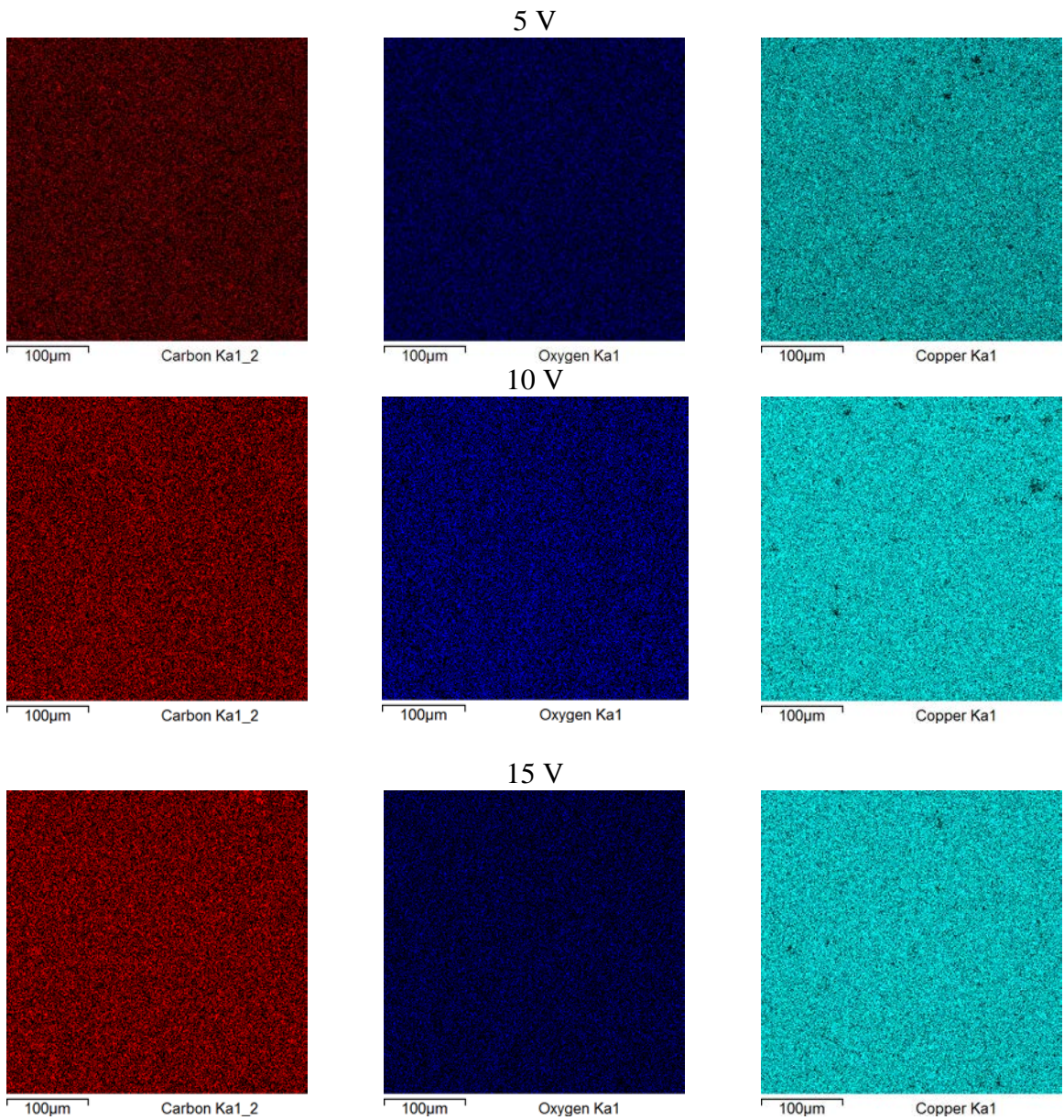

**Figure 4.** The EDX mapping of the Cu samples coated by GO at different voltages

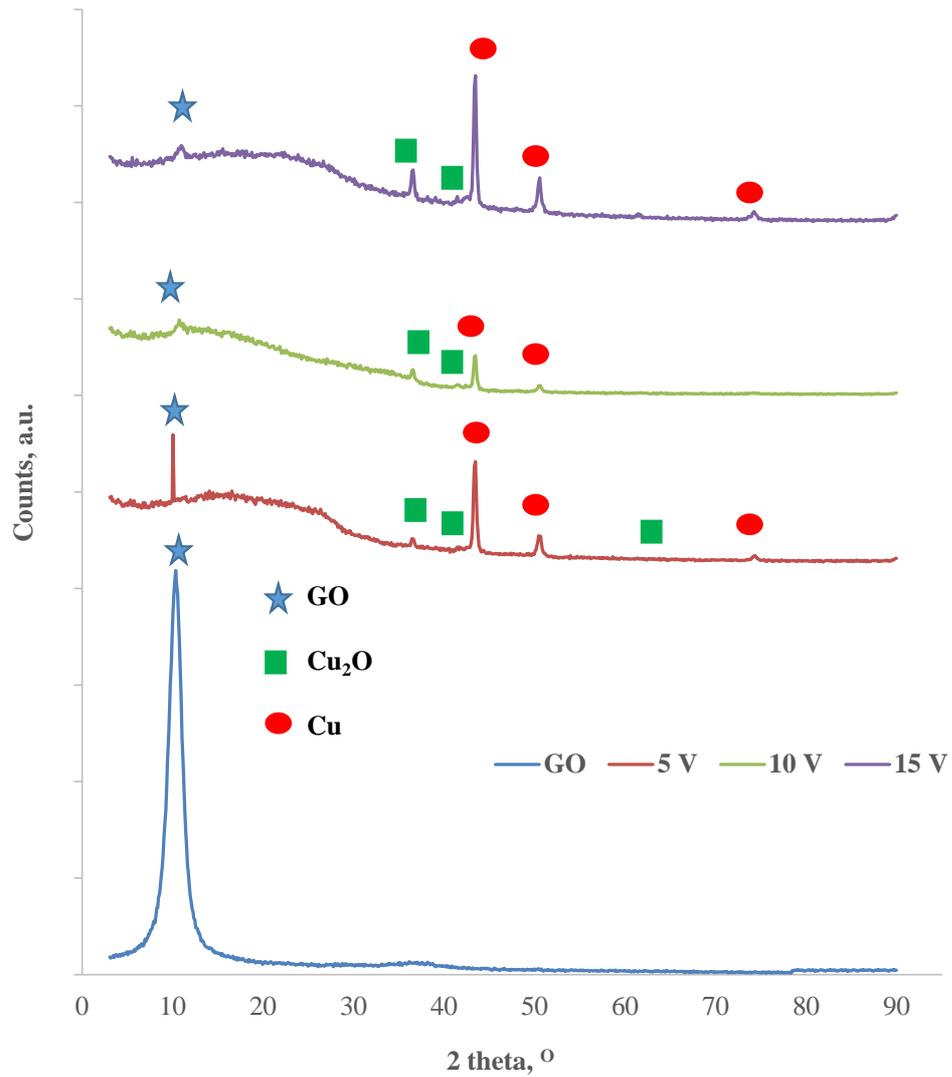

**Figure 5.** The TL-XRD patterns of the Cu samples coated by GO at different voltages.

The TL-XRD patterns of GO/Cu samples after EPD process reveal the distinct peaks corresponding to GO, copper and $Cu_2O$ (Fig. 5). The peak corresponding to graphene oxide is broadened concerning that of pure graphene oxide and height of this peak decreases with the value of applied current voltage. The broadening of graphene oxide peak on the samples after electrophoretic deposition may be caused by the formation of an amorphous phase and partial reduction of graphene oxide [40,41].

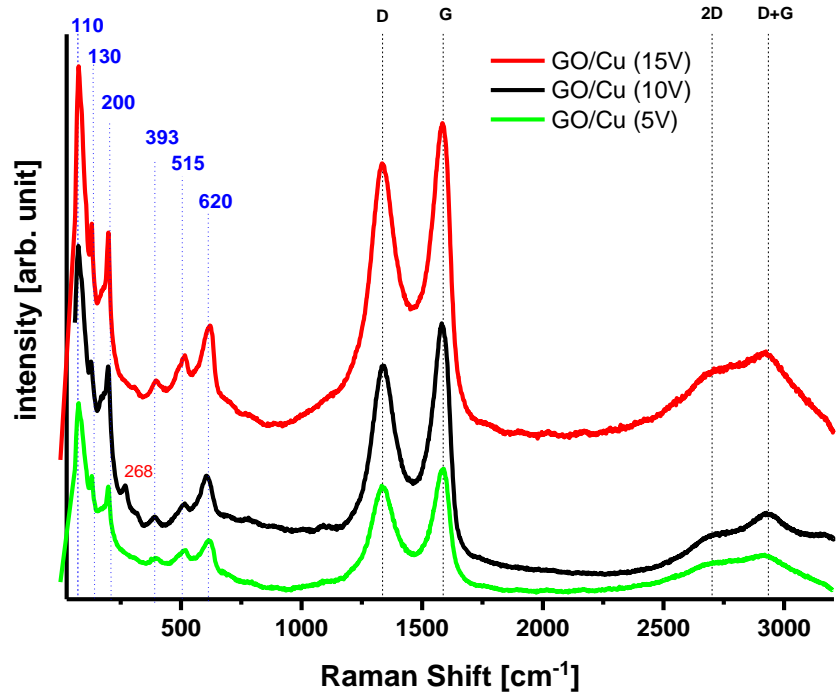

**Figure 6.** The Raman spectra of copper samples coated by GO at different voltages

*3.2. Raman spectra*

On Figure 6 the Raman spectra for copper samples prepared at different voltages are shown. Recorded spectra are a compilation of GO and $Cu_2O$ signals and are typical to all samples. The characteristic signals for GO materials, in that both exhibit the first-ordered G- and D-bands at around 1600 and 1350 $cm^{-1}$, respectively. The G-band is associated with *sp2*-hybridized carbon atoms and originates from the doubly degenerate zone center E2g mode. The D-band is correlated with *sp3*-hybridized carbon atoms as it requires a defect for its activation by double resonance, thus indicating the presence of lattice defects and distortions [42]. In addition, the spectrum has low intensity bands at 2680 $cm^{-1}$ and 2931 $cm^{-1}$, which correspond to the 2D and D+G bands. The 2D band is sensitive to the thickness of graphene and the ratio $I_{2D} / I_G$ makes possible to estimate the thickness of the GO layer. In the case of the tested materials, the ratio is 0.035 (5 V), 0.1 (10 V), 0.4 (15 V) which suggests a few-layered texture of GO on the analyzed samples.

The ratio of the D- to G-band intensities ($I_D/I_G$) is indicative of the quality of the graphitic lattice. This ratios are equal to 0.80 for the 10V and 5V sample and 0.88 for the 15V sample and are characteristic of GO. For the GO/Cu (5V) and GO/Cu (15 V) samples, prominent Raman peaks were observed at 110, 130, 200, 393, 515, and 620 $cm^{-1}$ corresponding to signals of $Cu_2O$. Based on literature reports, the observed Raman bands are assigned as follows: Raman allowed mode 3Γ_25(F2g) at 515 $cm^{-1}$, two IR active modes F1u(Γ15) at 150 and 630 $cm^{-1}$, and overtone

2Eu (2Γ12) at 200 cm$^{-1}$. The mode at 415 cm$^{-1}$ is due to a multiphonon process. The IR active bands at 110 and 130 cm$^{-1}$ are observed because of the violation of selection rules, which demonstrates the defects present in Cu$_2$O [43]. The Raman spectrum of GO/Cu (10 V) appears to be similar to that observed for the above two samples. However, in addition to the prominent peaks observed for Cu$_2$O and GO, a weak bands are also observed at 269 cm$^{-1}$ and 298$^{-1}$ (not marked) which corresponds to the Ag mode of CuO [44, 45]. The Raman scattering investigations of the investigated samples show that the morphology and size of structures have a great effect on the intensity of the two-phonon scattering band (2Eu) Raman spectra of these samples show that the grain size and crystallinity decreases with voltage, the Raman peaks shift to lower wavenumber and become broader owing to the size effects.

Raman mapping results of the Cu$_2$O 620 cm$^{-1}$ band and graphene G are shown in Fig. 7, respectively. Comparing the Raman mapping and the optical image of the graphene island, we find a positive correlation between the Raman intensity of Cu$_2$O and that of graphene. A positive correlation between the Raman intensity of Cu$_2$O and the D band shifts in graphene was also observed in Raman mapping (see Fig. 7).

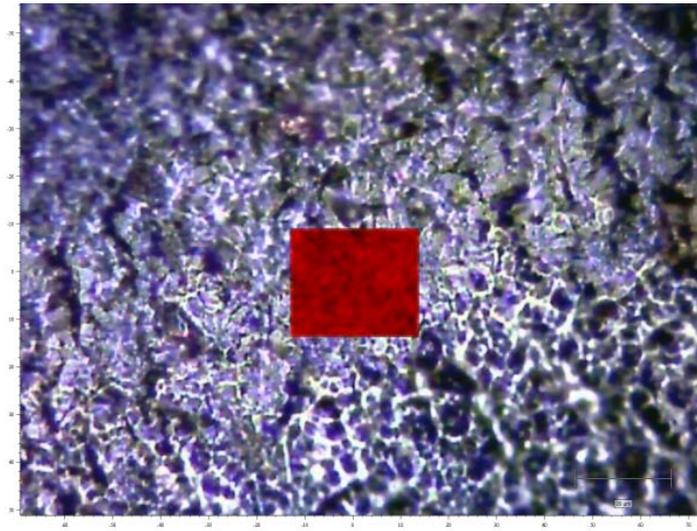

**Figure 7.** Exemplary optical image and combined Raman mapping image obtained at 620 cm$^{-1}$ Cu$_2$O (red area) and D band black area (sample prepared at 10 V)

Raman imaging with the Raman imaging microscope was used to identify defective regions in the graphene layer via spatially mapping the D band (1345 cm$^{-1}$) intensity to baseline ratio (as mentioned in the and also Cu$_2$O (620 cm$^{-1}$)), hence enabling us to correlate the Cu$_2$O to the GO distribution on the sample surface. GO is located on the surface (red area) of sample while Cu$_2$O is deeply inside the structure (dark areas).

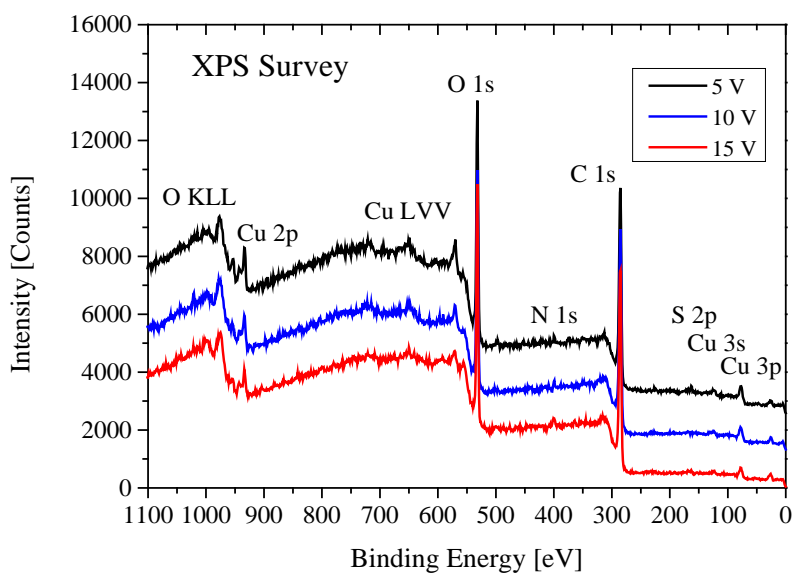

**Figure 8.** XPS survey spectra of GO/Cu interfaces

*3.3. X-ray photoelectron spectra*

On Fig. 8 are shown XPS survey spectra measured at 1100-0 eV binding energies of GO/Cu samples prepared by electrophoretic deposition at different voltages (5-15 V). One can see that XPS survey spectra show the only small content of nitrogen and sulfur impurities. The surface composition (in at.%) estimated from these spectra (Table 1) shows C/O ratio of 2.38-2.57. Which is found to be intermediate between that of GO (1.68) and rGO (8.32) [46]. Can it indicate the partial reduction of GO achieved during EPD-process or formation of Cu-O bonds? To receive an answer to this question we have measured high-energy resolved XPS C 1s-spectra of our rGO/Cu samples and compared them with spectra of GO [17] and rGO [47] in Fig. 9. One can see (Fig. 9 and Table 2) that contribution of C-O bonding with respect to C-C bonding is dominating (1.37) in GO and minimal in rGO (0.25) whereas in rGO/Cu composites it is not negligible (0.75-0.84) which definitely indicates for not full reduction of graphene oxide in our samples. The high-energy resolved XPS Cu 2p-spectra of rGO/Cu samples (Fig. 10) show a chemical shift concerning $Cu_2O$ and Cu-metal and appearance of charge transfer satellites ($S_1$ and $S_2$) typical for CuO which evidences a formation of CuO layer on the Cu substrate. Note, that the contribution from oxidized copper in the spectra changes insignificantly with the grown of the number of GO layers over the metallic substrate. Therefore oxidation of the substrate limited not by the accessibility of the oxygen from GO layers which contain enough oxygen-contained groups for further oxidation. We can speculate that observed self-limited oxidation of copper substrate is an intrinsic property of GO/Cu interface limited by the holistic structure of this system.

**Table 1.** The surface composition of GO/Cu samples (in at.%) and estimated C/O ratio.

| Sample | C 1s | O 1s | N 1s | Cu 2p$_{3/2}$ | S 2p | C/O |
|---|---|---|---|---|---|---|
| rGO/Cu (5 V) | 68.1 | 28.6 | 1.3 | 1.1 | 0.9 | 2.38 |
| rGO/Cu (10 V) | 70.9 | 27.5 | 1.2 | 0.4 | 0.1 | 2.57 |
| rGO/Cu (15 V) | 68.5 | 28.0 | 2.0 | 1.0 | 0.5 | 2.44 |

Based on obtained from XPS specters C:O ratio we can evaluate the thickness of deposited GO films. For this propose, we summarized atomic masses of carbon and oxygen atoms per area for given C:O ratio and divided it by the square of this area. The estimated weight of one partially reduced GO monolayer per cm$^2$ is $10^{-4}$ mg. Therefore increasing of the mass of the film at 0.1 mg/cm$^2$ is corresponding with deposition of about thousand of layers of GO. Taking into account that interlayer distance in GO varies from 12 to 6 Å [3-5] we could propose further estimation: increasing of the mass at 0.1 mg/cm$^2$ is matching with increasing of thickness about 90 nm. Thus the thickness of the films deposited at 5, 10 and 15 V (see Fig. 2) is in order of 490, 710 and 880 nm respectively.

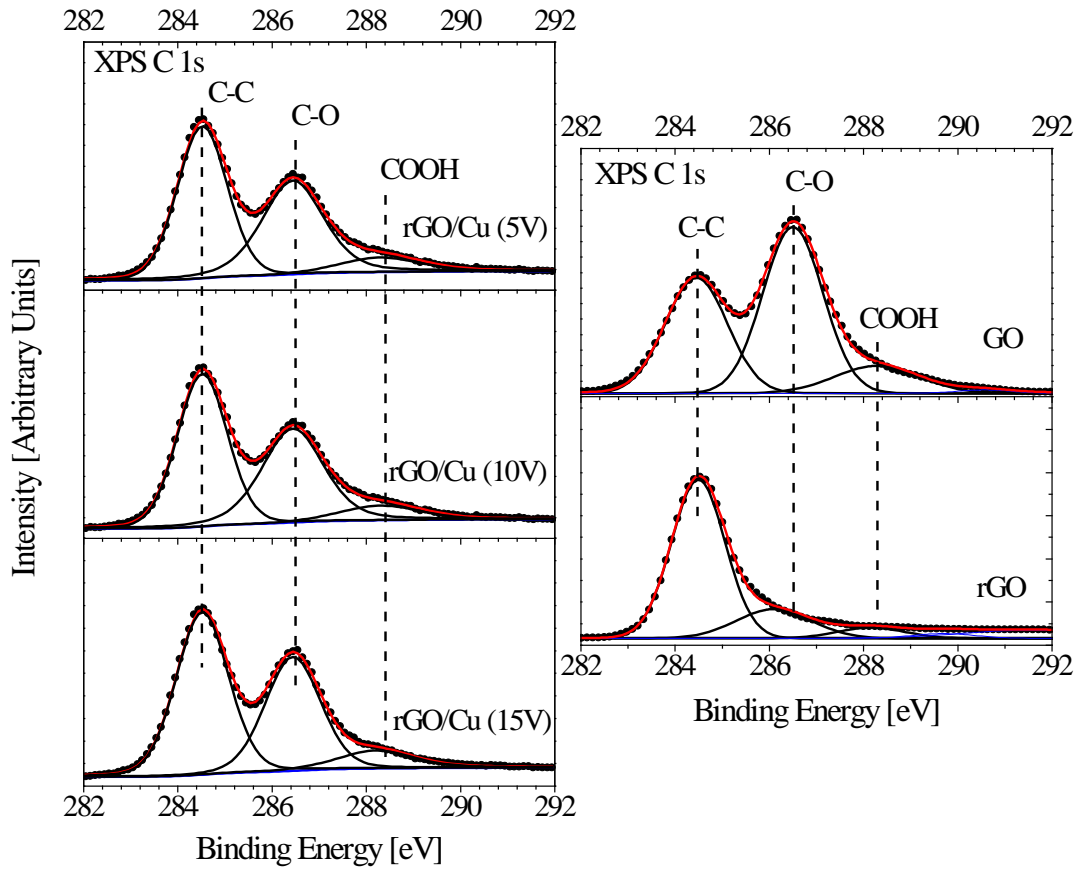

**Figure 9.** XPS C 1s spectra of GO/Cu samples (left panel) and reference samples (GO [17] and rGO [47] (right panel)).

**Table 2**. Contributions of C-C, C-O and COOH bonds to XPS C 1s spectra of rGO/Cu composites and reference samples.

| Sample | C–C | C–O | COOH | C-O/C-C |
|---|---|---|---|---|
| rGO/Cu (5 V) | 0.5 | 0.42 | 0.08 | 0.84 |
| rGO/Cu (10 V) | 0.5 | 0.42 | 0.08 | 0.84 |
| rGO/Cu (15 V) | 0.52 | 0.39 | 0.09 | 0.75 |
| rGO [17] | 0.79 | 0.2 | 0.01 | 0.25 |
| GO [46] | 0.37 | 0.51 | 0.12 | 1.37 |

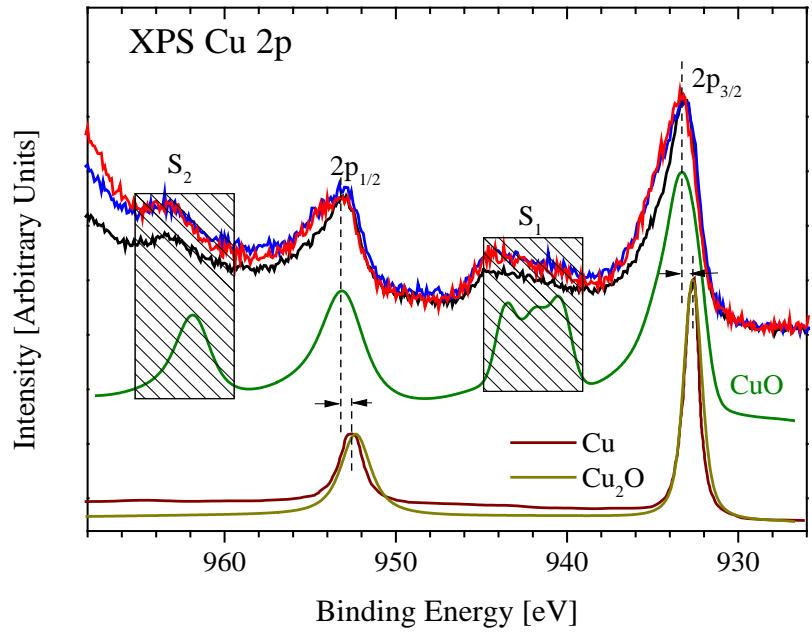

**Figure 10.** XPS Cu 2p-spectra of GO/Cu samples and reference samples (CuO, Cu$_2$O and Cu) [48]

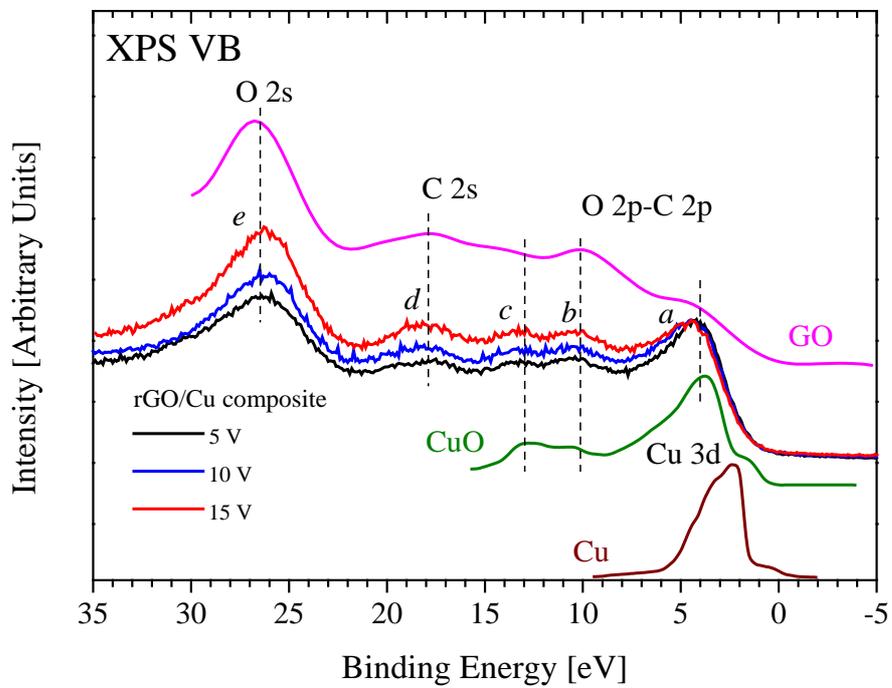

**Figure 11.** XPS VB spectra of GO/Cu and reference samples (rGO [49], CuO [50] and Cu [52]).

XPS VB spectra of GO/Cu samples (Fig. 11) reveal a very rich fine structure (*a-e*) which is similar to that of GO and CuO. The comparison with spectra of reference samples shows that it can be attributed to the superposition of that of partially reduced GO [49] and CuO [50]. Due to the closeness of the closest to Fermi level peaks of metallic Cu, copper suboxides on the surface ($Cu_xO$) [51] and bulk Cu [52] is rather difficult to identify peak an on VB with only one from discussed copper compounds.

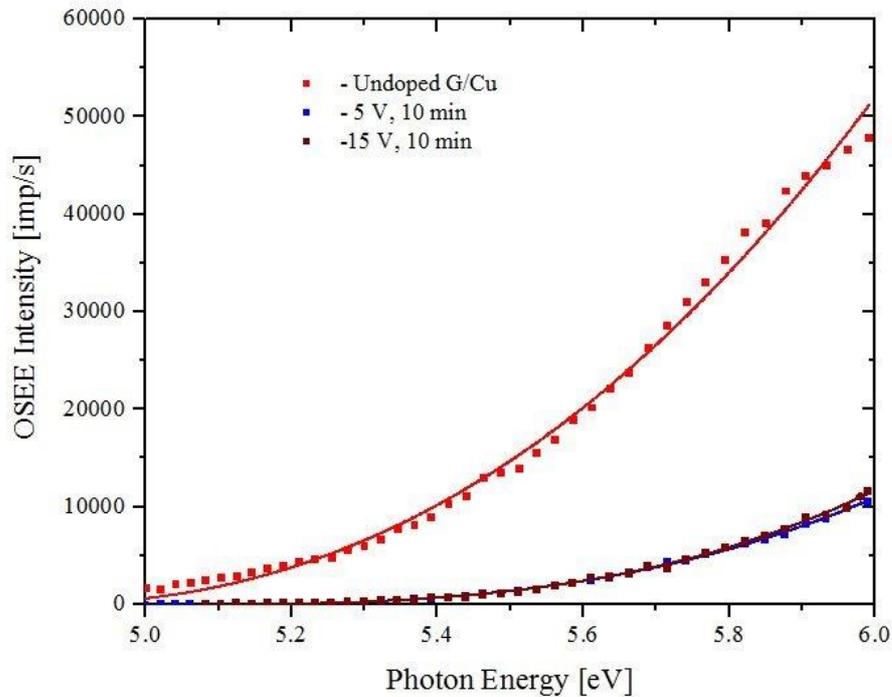

**Figure 12.** The OSEE spectra from GO/Cu samples (approximation is made for *n* = 2.5).

*3.4. Optically stimulated electron emission*

Because the GO/Cu samples have no luminescent activity (probably because of the small film thickness of GO) we have measured optically stimulated electron emission (OSEE). The experimental realization and information possibilities of this method for studying the characteristics of the energy band structure in case of the thin-film structures were demonstrated in our recent paper [53]. It should be noted that in the case OSEE from materials with bandgaps (e.g., insulators and semiconductors), the depth of the layer being analyzed can be from a few angstroms to more than 100 nm [54]. The dependence of the intensity of electron emission with photon energy $I = f(hv)$ is shown in Fig. 12. For the analysis of OSEE spectra we have used Kane's relation, which allows determining the number of parameters for the photoemission of samples under study [55]:

$$I = A(h\nu - \varphi)^n \qquad (1)$$

where, A - parameter approximation (scaling factor); hv – photon energy; φ - work function; n - coefficient characterizing the type of interband transitions (VB → CB) and it equals to 1.0; 1.5; 2.0; 2.5, depending on the prevalence in OSEE direct or indirect optical transitions.

**Table 3.** The results of analyzing for OSEE spectra of GO/Cu samples.

| GO/Cu (5 V, 10 min) | | GO (15 V, 10 min) | |
|---|---|---|---|
| $\varphi$, eV | $n$ | $\varphi$, eV | $n$ |
| 5.1 | 2.5 | 5.1 | 2.5 |

Table 3 shows the values of the parameters φ and n, which have been found in the approximation of the experimental dependence of the function (1). In agreements with previous results increasing the thickness of GO cover does not lead to visible changes in optical properties of GO/Cu composites. Results reported in Table 3 demonstrate that when n = 2.5 eV the work function of the investigated samples φ = 5.0-5.1 eV differ significantly from its value for the as-prepared GO (φ ~ 4.5 eV) but agree quite well with the electron work function for copper oxides (φ = 5.1-5.34 eV [55]). Basing on OSEE measurements we can conclude that the approximation of curves I = f(hv) by Kane's equation [55], which is commonly used to describe the emission properties of semiconductor surfaces, indirectly confirms the formation of copper oxides sublayers (which are the wide-gap semiconductors) due to interaction of oxygen atoms from not completely reduced graphene oxide with copper substrate. The value of n = 2.5 is characteristic of indirect optical transitions, indicating that the part of the phonon subsystem in the optical excitation and subsequent ionization of interface electronic states of oxidized copper layers. In other words, the opportunity of simultaneous implementation for various electronic transitions in OSEE confirms the assumption that heterogeneity of the oxidized layers takes place. Therefore the difference of OSEE curves for GO (GO/Cu undoped) and for GO/Cu samples and the presence in the specters of composites only copper oxide component suggest for the possible vanishing of the energy gap in GO after formation of GO coatings.

*3.5. Density functional theory calculations*

For an understanding of the atomic and electronic structure of GO/Cu interface, we have performed the theoretical modeling of this system at various levels of oxidation of graphene. Because the experimental results demonstrate that increasing of GO layers does not provide valuable changes in the electronic structure and optical properties we have analyzed for simplification of calculations only one layer of GO over Cu (111) surface. Results of calculations exhibit the formation of bonds between oxygen contained groups and copper substrate (Fig. 13). To evaluate the robustness of these bonds we calculated the binding energies by the formula:

$E_{binding} = [E(GO/Cu) - (E(Cu) + E(GO))]/n$,

where $E(GO/Cu)$, $E(Cu)$, $E(GO)$ are the total energies of GO/Cu system, pure copper surface and GO with same C/O ratio, respectively and $n$ – number of formed C-O-Cu bonds.

**Table 4.** The binding energies between GO and copper substrate (in eV per oxygen contained group) and total energies required for migrations of the oxygen-containing groups from GO to the substrate (eV) for various levels C/O of GO oxidation.

| C/O | $E_{binding}$ | $E_{migration}$ (-OH) | $E_{migration}$ (-O-) |
|---|---|---|---|
| 4/3 | 0.648 | -0.321 | -0.988 |
| 2/1 | 0.887 | +0.038 | -0.113 |
| 4/1 | 1.331 | +0.063 | +0.032 |

Results of calculations (Table 4) demonstrate the increasing of binding energy per oxygen-containing group with C/O ratio. Note that binding energy per unit of surface square remains almost the same and about three times higher than that of graphene/Ni interface. [57] The lowest value of binding energy remains rather high and of the same order as binding energy of epoxy groups (-O-) in graphene oxide at same C/O ratio [58] which evidences stability of GO/Cu system. For evaluation of the possibility of reduction of GO by migration of oxygen-containing groups to substrate we compared the total energies of systems before and after migration of hydroxyl (-OH) or epoxy (-O-) groups to substrate with formation of Cu-O chemical bond instead Cu-O-C. The results of calculations (Table 4) show that at a high level of oxidation of GO (C/O ratio 4/3) the migration of these groups from GO to copper is energetically favorable. This migration increase C/O ratio and at the value 2 the energies of migrations became almost equal zero and further decreased of the oxygen content of GO makes the reduction of graphene energetically unfavorable. The obtained results are in agreement with experimentally observed partial reduction of GO until C/O ratio about 2 independently from the voltage applied during deposition of GO on copper. Based on obtained experimental and theoretical results we can speculate that in the process of reduction of GO in the presence of metals the quality of the metallic surface plays a crucial role because the larger number of imperfectness as in the case of nanoparticles leads to favorability of migration of oxygen-containing groups. [11-14]

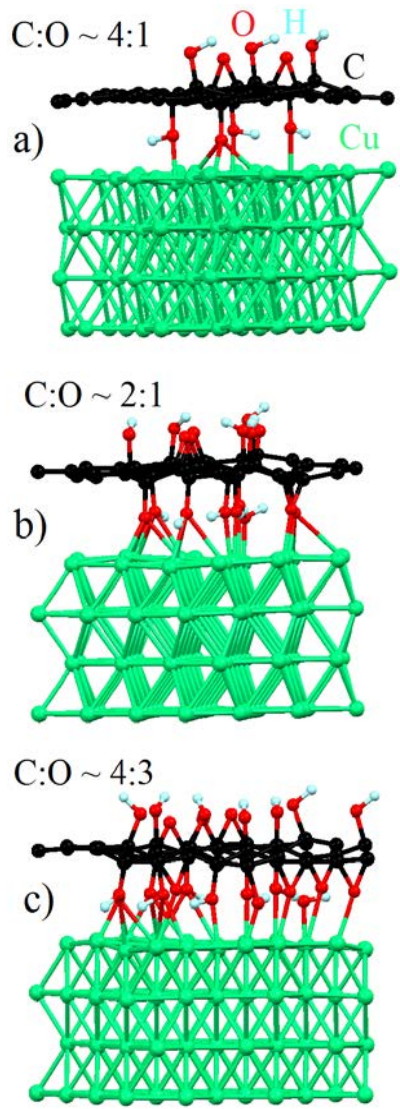

**Figure 13.** Optimized atomic structures of GO/Cu interfaces at various C/O ratio of GO.

The last step of our analysis is exploring of the effect of interface formation on the electronic structure of both GO and copper surface. Despite robustness of Cu-O-C bonds (Fig. 13 and Table 4), the formation of these bonds does not provide the significant changes in the electronic structure of a top layer of the copper surface which remains almost the same as for pure copper (Fig. 14a) almost for all values of C/O ratios of GO. For additional check we calculated densities of states of top layer for totally oxidized copper (111) surface (see inset on Fig. 14a) and found that its electronic structure close to XPS VBs of copper suboxides formed on early stages of the oxidation of copper surface [51] but different from the spectra of CuO (see Fig. 11). Also, the presence of metallic substrate under partially or oxidized upper layer provides the vanishing of the energy gap in this layer. Therefore the theoretical calculations demonstrate that for obtaining of CuO electronic structure (Fig. 11) and the opening of the energy gap in oxidized copper (see results of OSEE measurements) the oxidation of substrate beyond top layer is required. Because both theory and experiment evidence the limited reduction of GO we could conclude that some perforations in GO layers [59] provides a route for anodically generated and atmospheric oxygen to the metallic substrate.

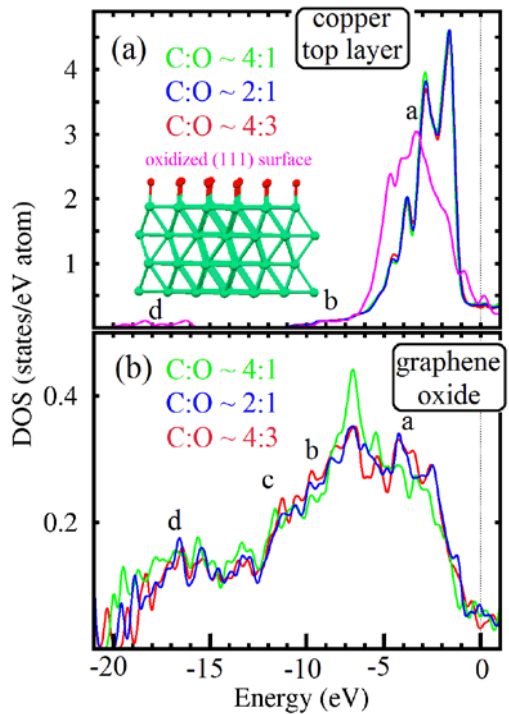

**Figure 14.** Densities of states of copper atoms from the top layer of copper substrate (a) and GO (b) for various C/O ratios in GO. On panel (a) is also provided optimized atomic structure (on inset) of totally oxidized (111) surface of copper and reciprocal density of states of copper atoms from the layer of this surface (violet line). The letters on the panels are corresponding with the marks of peaks in measured valence bands specters (Fig. 8).

The calculated densities of states of GO (Fig. 14b) are found to be in qualitative agreement with GO features observed in XPS valence band spectra (Fig. 11). The results of our calculations demonstrate that in contrast with free standing GO [58] the changes of C/O ratio do not provide the vavisible changes in the electronic structure of GO/Cu interface. Thus we can conclude that formation of robust (Table 4) Cu-O-C bonds and interaction with substrate significantly influence the electronic properties of GO. Another important effect of the interaction of GO with the substrate is the vanishing of the energy gap. This result is in qualitative agreement with OSEE measurements which do not detect the contribution from GO to the formation of energy gap (see Fig. 11 and Section 3.2). Thus we can conclude that in contrast with graphene/metal system where only light doping without the significant changes of electronic structure [58] is detected in the case of formation GO/metal interface the significant changes in the electronic structure of GO are found.

**Conclusions**

We have performed XPS, VB, OSEE measurements and DFT calculations of GO/Cu composite and found that formation of an interface between GO and copper substrate provides an only partial reduction of GO (with C-O/C-C ratio of 0.75-0.84). The fabrication of this composite provides a partial oxidation of copper surface with formation of several CuO-like layers between the copper substrate and GO. The level of reduction of GO and oxidation of metallic substrate does not depend on from the thickness of Cover. The theoretical modeling also demonstrates the formation of robust Cu-O-C bonds between GO and substrate. Formation of these bonds leads to significant changes in the electronic structure of GO simultaneously with negligible changes in the electronic structure of metallic substrate which detected in XPS VB spectra and obtained from DFT calculations. Both theoretical modeling and OSEE measurements also evidence vanishing of the band gap in GO after formation of GO/Cu interface.

**Acknowledgments**

The authors would like to acknowledge the FEI Company for the SEM characterization of GO coatings. XPS measurements were supported by the Russian Foundation for Basic Research (Project 17-02-00005), FASO (theme "Electron" № AAAA-A18-118020190098-5) and Government of Russian Federation (Act 211, agreement № 02.A03.21.0006). D. W. B. acknowledge support from the Ministry of Education and Science of the Russian Federation, Project №3.7372.2017/БЧ. This research was also funded by Rector's Grant in the field of research & development (Silesian University of Technology; grant No. 04/010/RGJ17/0050).